\documentstyle[aps,prl,multicol,epsfig]{revtex}
\begin{document}
\title{Interacting  Growth Walk on a honeycomb lattice}
\author{S. L. Narasimhan\cite{sln}
        P. S. R. Krishna\cite{sln}
        and M. Ramanadham\cite{sln}}

\address{\cite{sln}
                         Solid State Physics Division,
                         Bhabha Atomic Research Centre,
                         Mumbai 400 085, 
                         Maharashtra, India}
\author{K. P. N. Murthy\cite{kpn}
    and V. Sridhar\cite{kpn}}        
\address{ \cite{kpn}     Materials Science Division, 
                         Indira Gandhi Centre for Atomic Research,
                         Kalpakkam 603 102, 
                         Tamilnadu, India.}
\date{\today}
\maketitle
\begin{abstract}
The Interacting Growth Walk (IGW) is a  kinetic  algorithm
proposed recently for  generating long,  compact, self avoiding 
walks.  The growth process in  IGW is tuned by the
so called growth temperature $T' = 1/(k_B \beta ')$.
On a square lattice and at $T' = 0$,  IGW is
attrition free and hence  grows indefinitely.
In this paper we consider IGW on a honeycomb lattice. 
We take contact energy, see text, as $\epsilon=-|\epsilon|=-1$. 
We show that  IGW  at
$\beta' =\infty$ ($T'=0$)
is identical
to Interacting Self Avoiding Walk (ISAW) at 
$\beta=\ln 4$ ($k_B T = 1/\ln 4=0.7213$). Also  IGW at
$\beta ' = 0$ ($T' = \infty$) corresponds to ISAW at $\beta = \ln 2$
($k_B T= 1/ln 2 = 1.4427$).
For other temperatures we need to  introduce a 
statistical  weight factor to a walk  of the IGW ensemble to
make  correspondence with the ISAW ensemble.
\end{abstract}
\begin{multicols}{2}
We shall be concerned with a linear homo  polymer chain, modelled
by a lattice Self Avoiding Walk (SAW). Let $z$ denote the coordination
number of the lattice.
An ensemble of SAW is generated by the
following simple non-reversing blind ant
algorithm. A blind ant starts at a  site,
say origin. The ant steps
into one of the $z$  nearest neighbour sites with a probability
$p=1/z$. Since the ant never reverses its step, in  
the second and subsequent steps,
it moves to one of the $z-1$ nearest
neighbour sites with a probability
$(z-1)^{-1}$. If the ant finds that the site has 
been visited earlier, then the walk is terminated and
we start all over again. 
%When all the $z-1$ sites for the 
%next step are already visited, the walk can not proceed further;
Sample loss due to violation of self avoidance is called the problem 
of attrition because of which
growing of large number of long polymer chains becomes difficult. 
Thus, a  walk having
$N$ step is generated with a probability given by,
\begin{equation}\label{P_SAW}
{\cal P}_N ^{{\rm SAW}} ({\cal C}) =
 {{1}\over{z}}\left( {{1}\over{z-1}}
\right)^{N-1}.
\end{equation}
The important point is that as per the above non-reversing 
blind ant algorithm, all the $N$ step  
SAWs are generated
with the  same
probability.
Note that the only interaction present is due to the 
excluded volume effect (the hard core
repulsion). In the random walk model 
this is taken care of by the self avoidance condition.

Let us now
switch on a weak interaction. The aim is to model the interaction
that ensues when a segment of the polymer chain
comes close to another segment.
The segment - segment interaction can be attractive or repulsive
depending on the nature of the 
monomers present in the polymer chain.
This interaction is usually modelled, 
see for {\it e.g.} \cite{nbNN}  as follows. 
We say that in an SAW
configuration, every pair of occupied
nearest neighbour sites but not
adjacent along  the walk, carries an energy $\epsilon$.
We call such a pair,
as giving rise to a single non-bonded nearest
neighbour (nbNN) contact.
If $\epsilon < 0$ then the interaction is attractive and
if $\epsilon > 0$, it is
 repulsive. Thus a walk ${\cal C}$  belonging
to the SAW  ensemble
has an energy $E({\cal C}) = n_{NN}({\cal C})\times\epsilon$, where
$n_{NN}({\cal C})$ denotes the total number of nbNN contacts
present in  ${\cal C}$.

The probability with which
an SAW would be found in a canonical ensemble of 
Interacting Self Avoiding Walks (ISAW)
at temperature $T= 1/[k_B \beta]$ can now be expressed as,
\begin{eqnarray}
{\cal P}_N ^{{\rm ISAW}} ({\cal C}) &=&
 {{\exp \left[ -\beta E({\cal C})\right]}\over{
Q(\beta,N)}}.
\end{eqnarray}
In the above, the denominator is the 
canonical partition function, given by,
\begin{eqnarray}
Q(\beta,N) & = & \sum _{{\cal C}} \exp \left[ - \beta E({\cal C})\right],
\end{eqnarray}
from which
all the required macroscopic properties of the polymers can be
calculated either analytically or numerically,
in the thermodynamic limit of
$N\to\infty$. In what follows, we consider only 
attractive interaction, representing weak Van der Walls forces
and set $\epsilon = -|\epsilon|$, see for {\it e.g.} \cite{ab}
and without loss of generality take the
strength of interaction as unity, {\it i.e.} $| \epsilon | = 1$.

At very high temperatures,  $T\to\infty$ ($\beta\to 0$) the nbNN contact
interactions are  unimportant; a polymer configuration  is completely
determined by the excluded volume repulsion (present always 
and taken care of
by the self avoidance condition) and entropy.
We get a relatively
extended configuration representing a polymer under
good solvent conditions \cite{classics}. As $T$
is lowered, the
 contact interactions become more and more important and below a critical
temperature, called the theta point, there is an abrupt transition to
a collapsed phase. At the theta point itself, we get an intermediate
phase called the theta polymer.

A major problem  with SAW
is attrition, because of which we are not able to grow
large number of long polymer configurations within meaningful
computer times.
Several algorithms have been proposed and investigated
addressing the problem of attrition.
These include the True Self Avoiding Walk(TSAW)\cite{tsaw},
Kinetic Growth Walk(KGW) \cite{kgw},
Smart Kinetic Walk(SKW) \cite{skw},
Interacting Oriented Self Avoiding Walk (IOSAW) \cite{iosaw}
{\it etc.}, to name a few.  The most recent addition to this list
is the Interacting Growth Walk (IGW)
proposed by Narasimhan {\it et al} \cite{igw}. 
%%%%%%%%%%%%%%%%%%%%%%%%%%%%%%%%%%%%%%%%%%%%%%%%%%%%%%%%%%%%%%%%%%%%%%%%%%%%%%%%%%%%%%%%%%%%%%
\begin{figure}[htb]
{\samepage\columnwidth20pc
\centering
\epsfxsize=8.4cm
\epsfysize=7.4cm
\epsfbox{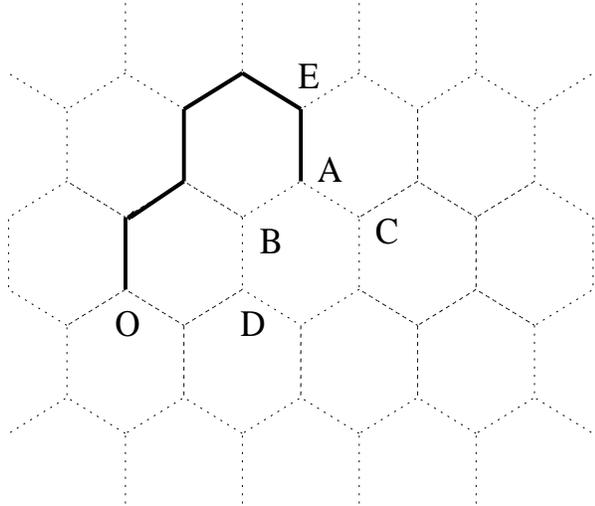}
\caption{\protect\small IGW on a honecomb lattice. The first step from
$O$ is taken with a probability $1/3$. All the subsequent
steps until site A are taken each with a probability $1/2$. If the
walk at site A goes to site B, then an nbNN contact
develops lowering the energy. The probability for this step
is $p(A\to B)=\exp (\beta')/[1+\exp(\beta')]$
On the other hand the step from A to C does not
change the energy and $p(A\to C) = 1/ [1+\exp(\beta')]$. If the
walk goes from A to B, then it is immediately followed by a step from
B to D with unit probability. 
                         }
\label{fig1_ps}}
\end{figure}
%%%%%%%%%%%%%%%%%%%%%%%%%%%%%%%%%%%%%%%%%%%%%%%%%%%%%%%%%%%%%%%%%%%%%%%%%%%%%%%%%%%%%%%%%%%%%%
In IGW, we define a growth temperature $T'$ that controls locally the
growth process as described briefly below.
We consider a non reversing and  myopic ant  rather than a blind ant.
The probability of moving
to a site that makes, say
$\mu$ number of nbNN contacts is made proportional to $\exp (\beta ' \mu)$.
This is illustrated in Fig. 1, where we consider SAW on a
honeycomb lattice.  A typical SAW trail is marked by thick
connected line segments in Fig. 1. For the honeycomb
lattice $z=3$ and $\mu = 0,\ 1$. The walk at site marked A in Fig. 1,
can move  either to the site B or to the site C.
 If the site B is selected, the energy
would be reduced by one unit since this choice
results  in an
nbNN contact. Hence we choose the site B  with a
relatively higher probability
determined by the growth temperature $T'$:
$p(A\to B) = \exp(\beta')/[1+\exp (\beta')]$.
On the other hand, the step from  site
A to site C, does not lead to any change in the energy, and
 $p(A\to C)=
1/\left[ 1+ \exp (\beta')\right]$.
Note that whenever
a contact making site is selected, it is followed  immediately
by a step to the
only remaining nearest neighbour site, and the probability
for this is unity. In Fig. 1, if the walk reaches site B then
then in its next step it goes to site D with a probability
 $p(B\to D) =1$. 
Let us say that in an $N$ step SAW, grown as per the myopic ant
IGW algorithm, there
are $n_{NN}$ contact steps accepted and $n' _{NN}$
contact steps avoided during the growth process. For example
in Fig. 1,. if the step A to B is taken  then we say the contact is
accepted. On the
other hand if the step A to C is taken then we say the
contact is avoided. We then have,
\begin{eqnarray}\label{P_IGW}
{\cal P} _N ^{{\rm IGW}}(\beta') &=&
{{1}\over{3}}\left( {{1}\over{2}}\right)^{N-1-2n_{NN}({\cal C})
 - n'_{NN}({\cal C})}\nonumber\\
&\times&
\left[  {{\exp(\beta')}\over{1+\exp(\beta')}}\right]^{n_{NN}({\cal C})}
\nonumber\\
&\times&
\left[ {{1}\over{ 1+\exp(\beta')}}\right]^{n'_{NN}({\cal C})}
\end{eqnarray}
Let us consider two extreme cases.\\

\noindent
Case 1 : $\beta '\to\infty$ ( $T'\to 0$)\\

For this case $n' _{NN}({\cal C}) = 0\ \forall\  {\cal C}$,
since whenever a contact making step
is available, the random walk takes it with probability unity.
In Fig. 1,
this corresponds to $p(A\to B) =1$ and $p( A\to C)=0$. Identifying
$E({\cal C})=-n_{NN}$, we have,
\begin{eqnarray}
{\cal P} _N ^{{\rm IGW}}(\beta'=\infty)= \exp
\left[ - (\ln 4) E({\cal C})\right]
\times {{1}\over{3}}\left( {{1}\over{2}}\right)^{N-1}
\end{eqnarray}
We immediately see
that the IGW ensemble at growth temperature $T' \to 0'$
($\beta '\to \infty)$
corresponds to ISAW ensemble at 
$k_B T = 1/\ln 4 = 0.7213$ ($\beta = \ln 4$).\\

\noindent
Case 2 : $\beta'=0$ ($T=\infty$)\\

At very high growth temperatures ($T'\to\infty)$, 
the nbNN contact interaction is unimportant
and the walk steps into one of the unvisited nearest neighbour
sites with equal
probability. We have,
\begin{eqnarray}
{\cal P}_{N}^{{\rm IGW}}(\beta'=0) =\exp \left[ - (\ln 2) E({\cal C})\right]
\times {{1}\over{3}}\left( {{1}\over{2}}\right)^{N-1}
\end{eqnarray}
Thus at very high growth temperatures ($T'\to\infty)$ ($\beta' \to = 0$)
IGW ensemble
is equivalent to ISAW ensemble at 
$k_B T = 1/\ln 2 = 1.4427$ ($\beta = \ln 2$).

Let us now investigate what happens when $0 < \beta' < \infty$.
We  rewrite Eq. (\ref{P_IGW}) as,
\begin{eqnarray}
{\cal P} _N ^{{\rm IGW}}(\beta')&=&
\exp\left[- \ln\left( F(\beta')\right) E({\cal C})
\right]\nonumber\\
&\times& \left[ {{2}\over{ 1+\exp(\beta')}}\right]^{n'_{NN}({\cal C})}
\times {{1}\over{3}}\left( {{1}\over{2}}\right)^{N-1} ,
\end{eqnarray}
where,
\begin{eqnarray}\label{fbetaprime}
F(\beta')&=&  {{4\exp (\beta')}
\over{1+\exp(\beta')}}
\end{eqnarray}
The bias arising  due to the contact sites avoided during the
growth process can be removed by attaching a weight $W(\beta',\ {\cal C})$
to a walk ${\cal C}$ generated as per the IGW algorithm. It is given by,
\begin{eqnarray}
W(\beta',{\cal C}) = \left( {{ 1+\exp(\beta')}\over{2}}\right) ^{n_{NN}'
({\cal C})}
\end{eqnarray}
The weighting for the bias removal is implemented as follows. Start an IGW
with a weight $W=1$. Every time a contact is available but not taken by the
random walk, multiply the weight by a factor given by $[1+\exp (\beta')]/2$.
Let $W(\beta', {\cal C})$ be the weight  at the end of an
$N$-step walk. Thus the IGW ensemble at $\beta'$ defined by the 
set of weights  
$\{ W(\beta' , {\cal C})\}$ is equivalent to ISAW at  $\beta = \ln F(\beta')$,
where $F(\beta')$ is given by Eq. (\ref{fbetaprime}).  

Infact, if we accept a weighted IGW ensemble, we can choose the weights
appropriately (called importance sampling \cite{pg}) 
to make a correspondence with ISAW ensemble, such that
$\beta'$ of IGW is the same as $\beta$ of ISAW. This is 
carried out as follows.
Let,
\begin{eqnarray}
F_1 (\beta')
&=& {{4}\over{1+\exp(\beta')}}\nonumber\\
F_2(\beta')&=&
 {{2}\over{1+\exp(\beta')}} .
\end{eqnarray}
Then we have,
\begin{eqnarray}
{\cal P}_N ^{{\rm IGW}}(\beta')&=&
\exp\left[ -\beta' E({\cal C})\right]\nonumber\\
&\times&
[F_1 (\beta')]^{n_{NN}({\cal C})}\nonumber\\
&\times&
[F_2 (\beta')]^{n'_{NN}({\cal C})}\times {\cal P}_N ^{{\rm SAW}},
\end{eqnarray}
where $E({\cal C}) = -n_{NN}({\cal C})$ and
${\cal P}_N ^{{\rm SAW}}$ is given by Eq. (\ref{P_SAW}) with $z=3$.
The factors which depend on $F_1$ and $F_2$ can be taken care of by
attaching suitable statistical weights to the configuration.
Essentially, 
we start the random walk with a statistical weight $W=1$.
At any stage of the growth process, if both the nearest neighbour
sites are  not contact making sites, then
we choose one of them with equal probability and proceed. We do not
do anything to $W$. In Fig. 1, when the walk goes from
the site E to A, we do not adjust  $W$.  On the other hand
if the step leads to  a contact (the step A to B in Fig. 1),
 then
$W$ is multiplied by
$1/F_1(\beta')$.
If the step avoids a contact (A to C in Fig. 1),
then we multiply $W$ by
$1/F_2 (\beta')$. Note that these weight adjustments are 
carried out in the same spirit as the PERM B algorithm 
proposed by Grassberger \cite{pg}. We carry out the above
weight multiplication  everytime
a contact making step or a contact avoiding step
is made during the growth process.
Let $W(\beta', {\cal C})$ be the weight at the end 
the $N$ step walk ${\cal C}$.
It is immediately seen that if the IGW is
weighted as per the above procedure, then
the growth temperature $T'$ is the same as the temperature  $T$
of the
canonical ensemble of ISAW.
Thus we have an IGW ensemble defined by the weights $\{ W({\cal C}, \beta)\}$.
We can employ this ensemble to calculate the macroscopic properties
like energy, specific heat (fluctuations in energy),
end - to -  end distance {\it etc.}

In conclusion, we have shown that IGW at $\beta'=\infty$
is identically equivalent to ISAW at $\beta = \ln 4$.
Also IGW at $\beta'=0$ is identically equivalent to
ISAW at $\beta = \ln 2$. For other values of $\beta'$ we introduce
a weight factor to correct for the bias arising due to the
contacts avoided during the growth process and show that the
weighted IGW at $\beta '$ is equivalent to ISAW at $\beta$ given by
$\ln F(\beta')$, where $F(\beta')$ is  given Eq. (\ref{fbetaprime}). 
Infact by choosing appropriate weights (importance sampling) 
for the accepted as well as avoided contacts we can render
weighted IGW at $\beta'$ as equivalent to ISAW 
at $\beta = \beta'$.

\end{multicols}
\end{document}